\documentclass[letter, oldversion]{aa}
\usepackage[dvips]{graphicx}

\usepackage{rotating}
\usepackage{longtable}
\usepackage{amssymb,amsmath}
\newcommand{\be}{\begin{equation}}
\newcommand{\ee}{\end{equation}}
\usepackage{latexsym}
\newcommand{\simless}{\mathbin{\lower 3pt\hbox {$\rlap{\raise 5pt\hbox{$\char'074$}}\mathchar"7218$}}} 
\newcommand{\simgreat}{\mathbin{\lower 3pt\hbox
     {$\rlap{\raise 5pt\hbox{$\char'076$}}\mathchar"7218$}}} 

\newcommand{\cmq}{cm$^{-3}$}

\newcommand{\sori}{$\sigma$~Ori}
\newcommand{\Msun}{M$_\odot$}
\newcommand{\Lsun}{L$_\odot$}
\newcommand{\Ha}{H$\alpha$}

\newcommand{\Lstar}{L$_{star}$}

\newcommand{\Mstar}{M$_{star}$}
\newcommand{\Lacc}{L$_{acc}$}
\newcommand{\Macc}{$\dot M_{acc}$}
\newcommand{\Mloss}{$\dot M_{loss}$}
\newcommand{\Mwind}{$\dot M_{wind}$}
\newcommand{\Ldisk}{L$_{disk}$}
\newcommand{\Myr}{M$_\odot$/yr}

\begin{document}

\title{ Discovery of an old photoevaporating disk in  $\sigma$~Ori
\thanks{
Based on observations collected at the European Southern Observatory, Chile.
Program 074.D-0136(A).}
}

\author{
E. Rigliaco\inst{1,2},
A. Natta\inst{1},
S. Randich \inst{1},
\and
G. Sacco \inst{3}
}

\institute{
    Osservatorio Astrofisico di Arcetri, INAF, Largo E.Fermi 5,
    I-50125 Firenze, Italy
\and
Universit\`a  di Firenze, Dipartimento di Astronomia, Largo E.Fermi 2,
    I-50125 Firenze, Italy
\and
Osservatorio Astronomico di Palermo, INAF, Piazza del Parlamento 1, 90134 Palermo, Italy
}

\offprints{erigliaco@arcetri.astro.it}
\date{Received 17 December 2008; accepted 02 February 2009}

\authorrunning{Rigliaco et al.}
\titlerunning{SO587}

\abstract
{The photoevaporation of circumstellar disks is a powerful process in the 
disk dissipation 
at the origin of the Orion proplyds. 
This Letter reports the first detection of a photoevaporating disk
in the final but long-lasting phase of its evolution.
The disk is associated to a low-mass T Tauri member 
of the \sori\ Cluster. It is
characterized by a very low (if any) accretion rate
and by a tenuous (\Mloss $\sim 10^{-9}$ \Myr) photoevaporation
wind, which is unambiguously detected in the
optical spectrum of the object. The wind
emits strong forbidden lines of [SII] and [NII]
because the low-mass star is close to a
powerful source of ionizing photons,
the O9.5 star \sori.}


\maketitle

\section {Introduction}

The  evolution of
circumstellar disks surrounding low-mass T Tauri stars (TTS in the
following) is controlled by the interplay
of different physical processes, which include 
viscous accretion onto the central star, photoevaporation by the
stellar radiation field, and planet formation. 
If the low-mass star is sufficiently close to a more massive  and
hotter object, 
photoevaporation by this external source of high-energy radiation is also
important because it can dominate the disk evolution.

The process of photoevaporation has been extensively discussed 
(see, e.g. Hollenbach et al.~PPIV; Dullemond et al.~PPV). High-energy photons heat  gas disks to temperatures
such that the thermal pressure exceeds the gravity from the central star.
The disk evaporates from outside-in, with a mass-loss
rate that decreases with time as the disk shrinks. 
Most of the well-studied Orion proplyds 
are caught in the first, short evolutionary phase, when 
the mass-loss rate is very high ($\dot M_{loss}>10^{-7}$ \Myr, Henney et al.,~1999). 

We report in this Letter on the first detection of a proplyd in 
a much
later evolutionary phase, when the mass-loss rate is $\ll 10^{-8}$ \Myr.
The object (a low-mass TTS) is located in the star-forming region
\sori\ (age $\sim 2-3$ Myr), which  contains the massive quintuplet system \sori.
The brightest star of this
system (spectral type O9.5) forms a large, low-density HII
region (Habart et al.,~2005) and the bright PDR known as the Horsehead (Abergel et al.,~2003). 
The low background from the HII region and the relatively short distance ($\sim 400$ pc;
Mayne \& Naylor,~2008) allow us to detect the slow photoevaporated TTS wind and to measure its
optical line emission spectrum.

\section {The T Tauri star SO587}

\subsection {Stellar properties}

SO587 (also identified as R053833-0236 Wolk J.S.,~1996 or Mayrit 165257 Caballero J.A.,~2008)  
($\alpha_{2000}$=+05:38:34.04, $\delta_{2000}$=-02:36:37.33) has been classified as M3-M4 by
Zapatero-Osorio et al.~(2002) on the basis of low-resolution optical spectra 
($T_{eff}\sim 3300$K). The
optical extinction is negligible (Oliveira et al.,~2004), and we estimate
a luminosity of  0.3$\pm 0.1$ \Lsun, based on this spectral type and the
V, R, I magnitudes (Wolk, 1996).
The corresponding mass is about 0.2 \Msun, both from D'Antona \& Mazzitelli~(1997) and Baraffe et al.~(1998) evolutionary tracks.
Based on its location on the HR diagram,
SO587 has an  age of $\sim 1$ Myr, apparently slightly younger than the bulk of
the \sori\ stars.
SO587 is an X-ray source, with luminosity
$L_X\sim 10^{29} \,\, \rm{erg/s}$ 
(Franciosini et al.,~2006)

\begin{figure}[htbp]
\centering
\includegraphics[width=10cm]{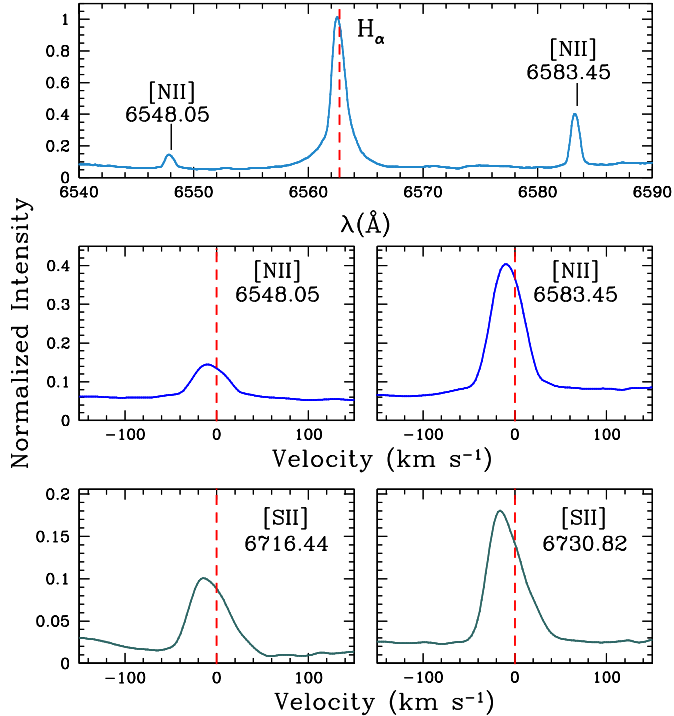}
\caption{The top panel shows the spectrum in the
\Ha\ region, with \Ha\ and the two [NII]$\lambda$6548.05 and [NII]$\lambda$6583.45 lines. Middle panel:
[NII]$\lambda$6548.05 and [NII]$\lambda$6583.45 line profiles as function of the
velocity shift with respect to the stellar velocity (vertical dashed line). Bottom panel:
same for [SII]$\lambda$6716.64 and [SII]$\lambda$6730.82, as label.
}
\label {lines}
\end{figure}

%

%

%

\subsection {Disk properties}

S0587 is detected by Spitzer at all IRAC wavelengths and at 24 $\rm\mu m$ with MIPS.
It shows  a relatively weak excess emission, and  is classified 
as an evolved disk (EV, Hernandez et al.~2007). We confirm this 
classification.
The IR emission is 
reproduced well by a geometrically flat, optically thick disk 
heated by the central star, seen at an inclination of about 40 deg. 
The outer disk radius is not constrained by the existing infrared
photometry, which is limited to 24 $\mu$m, as long as
$R_{out}\simgreat 1-2$ AU.
The model is not unique but  exploration of a large number of disk models
rules out  optically thin disks,
which have different
Spitzer colors and lower luminosity than observed in SO587 (\Ldisk/\Lstar$\sim 0.07$)
(see Cieza et al.,~2007).
Also, there is no evidence in the SED
of the large inner holes (few AUs) seen in transitional disks (Chiang \& Murray-Clay,~2007 and references therein), as all the
models that fit the data have inner radii $\simless$ 0.1 AU.


\begin{figure}
   \centering
   \resizebox{\hsize}{!}{\includegraphics[]{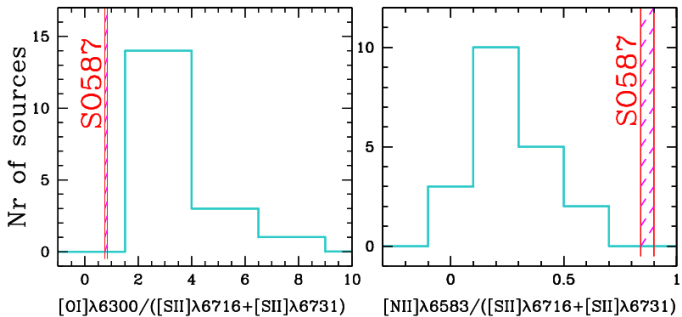}}
        \caption{ Distribution of the ratios [OI]6300/([SII]6716+[SII]6731) and [NII]6583/([SII]6716+[SII]6731)
among the Hartigan et al.~(1995) Taurus TTS with detected lines. The magenta dashed region shows the location 
of SO587 considering the errors on the line equivalent widths; [OI] from Zapatero-Osorio et al.,~2002.
}
\label {histogram}
\end{figure}

\subsection {Accretion rate }

SO587 shows no evidence of significant accretion onto the central star.
An estimate of \Macc\ can be obtained from the U band magnitude $18.5 \pm 0.8$ 
measured by Wolk (1996).
For the spectral type and luminosity of SO587, this corresponds to an excess 
U band emission of 1 mag at most. 
Using the correlation between the U band excess emission and the accretion luminosity
established by Gullbring et al.~(1998) for TTS, we derive 
\Lacc$\sim 3\times 10^{-3}$ \Lsun\ with an uncertainty of a factor 2, i.e., about 1\% of the stellar luminosity. 
Since a
comparable U band excess emission may be due to chromospheric activity (White \& Ghez, 2001), this is
likely  an upper limit to \Lacc.
The corresponding accretion rate is $\simless 3\times 10^{-10}$ \Myr.


 
\section { The optical spectrum}

SO587 was observed by Sacco et al. (2008)
using 
the multi-object instrument FLAMES on VLT/UT2 and Giraffe spectrograph
with the HR15N grating (6470-6790 \AA, spectral resolution R=17,000). 
The observations were obtained in 6 separate runs of 1 hour each 
in October and December 2004. 
Details on the observations and data reduction, in particular on 
sky subtraction, can be found in Sacco et al.~(2008). 
The spectra of SO587 show 
H$\alpha$ in emission, as well as strong forbidden lines from 
[NII] at 6548 and 6583 \AA\ and [SII] at 6716 and 6731 \AA\ 
(Fig.~\ref{lines}).
Table~\ref{peak} gives the pseudo-equivalent width (pEW) of each line 
(negative values for emission lines), the FWHM (not corrected
for the instrumental resolution of about 20 km/s), and the line luminosities.
The contribution 
to the lines (both \Ha\ and the forbidden lines) 
of the extended HII region due to the O9.5 star \sori\  is negligible.

The line luminosities were computed using the observed R magnitude to calibrate the
continuum flux and the wavelength dependence within the band  appropriate for a 3300K star.
The luminosity is of the order of $10^{-5}$ \Lsun\ for all the forbidden lines.

Zapatero-Osorio et al.~(2002) also detected 
\Ha\ and forbidden emission lines in two of their low-resolution spectra with highest quality, 
but not in the other, lower-quality spectra; 
when detected, the pEW of the lines are similar to our values.
We checked for evidence of line variability in the Sacco et al. spectra,
and found none.
The line profiles and intensities are very similar in each run,
with the pEW varying $\sim$ 10\%.

In the two spectra where emission lines were detected,
Zapatero-Osorio et al.~(2002) also measured the pEW of the
two oxygen lines at 6300 and 6364 \AA, which are outside 
our spectral range. 
The corresponding luminosity is $1.2 \pm 0.06 \times 10^{-5}$ \Lsun\ for
the 6300 \AA\ line and $0.4 \pm 0.03 \times 10^{-5}$ \Lsun\ for the
6364 \AA\ one, similar to  the [SII] and [NII]
line luminosity.


\section { Not an accretion-driven wind}

The forbidden lines detected in SO587 are seen in many TTS, and are usually interpreted in the 
framework of accretion-driven winds (e.g. Ferreira et al.,~2006 and reference therein).
These centrifugally-driven MHD winds originate in the inner disk and move fast. 
An estimate of the average mass-loss rate can be obtained from the
luminosity of the forbidden lines, following
Hartigan et al.~(1995). Taking, e.g., the 
[SII]$\lambda$6731 line,  this is
\begin {equation}
\begin{split}
\dot{M}_{wind} = 3.38 \times 10^{-8}\left( 1+\frac{n_c}{n_e}\right) \left( \frac{L_{6731}}{10^{-4}\rm{L_{\odot}}}\right)  \times \\
\left( \frac{V_{\perp}}{150 \, \rm{km \,\, s^{-1}}}\right) \left( \frac{l_{\perp}}{2 \times 10^{15}\, \rm{cm}}\right)^{-1}\,\, \rm{M_{\odot}/yr}.
\end{split}
\label{Mwind}
\end {equation}
The electron density $n_e$, derived from the ratio of the two [SII] lines, is $\sim 1.5 \times 10^{3}$ \cmq, well below the critical density $n_c$, which is 
$\sim 1.3 \times 10^{4}$ \cmq.
Assuming  that all S is SII, a S abundance of $1.6\times 10^{-5}$, a typical value of $V_{\perp} \sim 150 \, \rm{km\,\,s^{-1}}$ for the outflow
velocity and that the outflow fills
our beam of 1 arcsec ($l_{\perp} \sim 6\times 10^{15}$  cm), we derive a mass-loss rate of about $10^{-8}$ \Myr. 

This value of \Mwind\ is high for a 0.2 \Msun\ TTS (Hartigan et al., 1995). It is
particularly high when compared to \Macc, as \Mwind/\Macc$\simgreat$10.
In TTS, this ratio is $< 1$ and $< 0.1$ in the majority of cases 
 (Hartigan et al.,~1995; White \& Hillenbrand,~2004), in agreement with the expectations of accretion-driven jets and winds
(e.g., Shu et al.,~1994; Pudritz et al.,~2007). 
Even taking the uncertainties on the assumptions 
entering into Eq.(1) into account 
(see Cabrit,~2002), the 
very high value of the ratio \Mwind/\Macc\ makes it unlikely that the 
forbidden lines 
in SO587 could be  emitted by an accretion-driven wind.

\section {Line profiles}

Additional evidence that the  SO587 forbidden lines do not originate
in an accretion-driven wind is provided by a comparison of the forbidden line 
luminosities and profiles with those of classical TTS.
We first notice 
the strength of the  [NII] lines, which 
are  weak or absent in TTS,  but
comparable in luminosity to  the [SII] lines in SO587.
Strong [NII] lines are expected in highly ionized gas,
as in HII regions or in the bow 
shocks at the head of jets, and are weaker than the other forbidden lines in the partially neutral wind/jets of TTS. 
SO587 is also unusual in having strong [SII] lines  with respect to the
[OI] lines (Fig.~\ref{histogram}). 

Moreover, the profiles of the SO587 [SII] and [NII] lines differ from the typical TTS line profiles. 
The [SII] and [NII]  are resolved in our spectra, and they show a small shift of the peak to the blue ($\simless 10-14 \, \rm{km \,\, s^{-1}}$), 
a steep rise in the blue, and a slow decline in the 
red wing.
Spatially unresolved profiles of
the [SII] 6731 \AA\ line in a sample of accreting TTS
typically show larger peak blueshifts ($\sim 80-100 \, \rm{km \,\, s^{-1}}$),
and significantly more emission in the blue than in the red wing
(Edwards et al.~1987; Hartigan et al.~1995). In fact, 
these asymmetries are  interpreted as evidence
that the disk obscures the 
receding portion of the line emitting region. 

The high luminosity in [NII], the low ratio [OI]/[SII] $\sim 1$, and the
details of 
the line profiles of SO587 are strongly reminiscent
of the properties of the Orion  proplyds
(Bally et al.,~1998; Henney \& O'Dell,~1999). 
The Orion proplyds are interpreted as the 
result of photoevaporation
of the disk by the hot Trapezium stars
(see, e.g. Hollenbach et al.~PPIV). 
The optical forbidden lines of ionized species, such as [SII] and [NII],
are formed
in the outer part of the outflow, which is ionized, heated, and shocked by the
radiation and wind of ${\theta}^1C$ Ori (Garcia-Arredondo et al.,~2001).

\begin{table}[!t]
\caption{Characteristics of the emission lines.}

\begin{center}

\begin{tabular}{c c c c c c}

\hline \hline
line & $\lambda$ & pEW  & FWHM & Luminosity\\
& (\AA) & (\AA) &  ($\rm{km \,\, s^{-1}}$) & $10^{-5}$\Lsun \\
\hline
&&&&\\
\Ha\ & 6562.71 & -15.2 $\pm$ 0.1  & 69.9 $\pm$ 2 &  12.0  $\pm$ 0.1\\  
$[\rm{NII}]$ & 6548.05  & -0.82 $\pm$ 0.2 &  42.6 $\pm$ 1 & 0.6 $\pm$ 0.15\\
$[\rm{NII}]$ & 6583.45 & -2.46 $\pm$ 0.1 &  39.6 $\pm$ 0.5 &  1.9  $\pm$ 0.1\\
$[\rm{SII}]$ & 6716.44 & -1.22 $\pm$ 0.03 &  48.7 $\pm$ 0.4 & 0.6 $\pm$ 0.02\\
$[\rm{SII}]$ & 6730.82 & -1.60 $\pm$ 0.08  & 43.2 $\pm$ 0.4 & 0.8 $\pm$ 0.04\\
&&&&\\
\hline\hline

\end{tabular}

\end{center}
\label {peak}
\end{table}







\section {A photoevaporating disk}

We propose that 
SO587 is a photoevaporating disk, which we detect in optical forbidden lines
because it is ionized
by the nearby  (projected distance of $\sim$0.35 pc)  O9.5 star
\sori. 
Let us assume that the SO587 disk loses mass at a rate \Mloss. 
Whatever the source of the photoevaporating photons (the central star or \sori\ -- see below), 
this wind originates in the outer disk, and is slow (few $\rm{km \,\, s^{-1}}$) and neutral. However, 
the high-energy ($h\nu>$13.6eV) photons from \sori\
ionize the outflowing gas to a distance $r_{IF}$ from SO587 roughly

\begin{equation}
r_{IF} \sim  1.6 \times \Big({{\dot M_{loss}}\over{\mu_H v_{IF}}}\Big)^{2/3} \> \Big({{\alpha \Delta^2}\over{\Phi_i}}\Big)^{1/3}.
\label{intIF}
\end{equation}
Here $\alpha$ is the recombination coefficient ($\alpha = 2.6 \times 10^{-13} \, \rm{cm^{3} \, s^{-1}}$), 
$\mu_H$ the mean molecular weight 
$\Phi_i$ the high energy flux from \sori\ ($\Phi_i \sim 10^{48}$ photons/s; Peimbert \& Rayo,~1975), $\Delta$ is the distance between SO587 and \sori, and
$v_{IF}\sim 10 \, \rm{km \,\, s^{-1}}$ the velocity at the ionization front (St\"oerzer \& Hollenbach,~1999), so that 
\begin{equation}
r_{IF} \sim 80 {\rm AU} \> \Big({{\dot M_{loss}}\over{10^{-9} \rm{M_\odot / yr}}}\Big)^{2/3}\> \Big({{\Delta}\over{0.35\, \rm{pc}}}\Big)^{2/3}.
\label{intIF2}
\end{equation}

The ionization front moves away from SO587 and is closer to \sori, if 
either \Mloss\  and/or the distance between the two stars increases. 
The optical forbidden lines shown in Fig.~\ref{lines} can only come
from the ionized regions of the outflow; since they are detected
in spectra taken with a beam of radius $\sim 200$ AU
centered on SO587,
it must be $r_{IF}\ll 200$ AU. This condition is verified
for $\dot M_{loss}\simless 10^{-9}$ \Myr; if \Mloss\ is significantly higher,
$r_{IF}$ moves too far from SO587, and the outflow region within
the beam is all neutral. 

In this model, the [SII] and [NII] forbidden lines originate
mostly in the ionized gas at $r_{IF}$. For \Mloss=$10^{-9}$ \Myr, the
electron density at $r_{IF}$ from the continuity equation 
($n_e= \dot M_{loss}/(4\pi r_{IF}^2 \mu_H v_{IF}$)
is $\sim 2 \times 10^3$ \cmq, in good agreement
with the electron density derived from the observed ratio of the two [SII] lines ($\sim 1.5 \times 10^3$ \cmq; \S 4).
The [SII] 6731 \AA\ line luminosity within the beam, computed
assuming that
all S in SII and a temperature $\sim 10^4$K, 
is $\sim 10^{-5}$ \Lsun, in good agreement with the observations. \\

What is the origin of this outflow?
Both the SO587 star and \sori\ can, in principle, cause disk
photoevaporation, and the properties of the outflow are, qualitatively, similar.
Let us consider first the case of  disk photoevaporation 
due to the FUV (6eV$<h\nu<$13.6eV) radiation of \sori.
Adams et al.~(2004) have computed detailed models of the photoevaporation 
by an external source
of disks around stars of different masses.
At a distance of 0.35 pc, the FUV radiation field from \sori\
is $G_0\sim 10^4$ (in units of the Habing radiation field; Abergel et al.,~2003).
Let us assume that the SO587  disk 
has mass $\sim 0.05$ \Mstar\ when the outer radius is $\sim$ 60 AU  and
surface density $\propto R^{-1}$. 
If we take Adams et al.~(2004)  results for a star
of central mass 0.25 \Msun\ in a radiation field $G_0=3000$
(the exact value of $G_0$ is not crucial as long as it is higher than this
value),  the disk has 
a first, evolutionary stage characterized by a relatively
large size and high mass-loss rates
$\sim few \times 10^{-8}$ \Myr. 
This phase is rapid, and the disk  loses about 80\% of its mass in $\sim 3\times 10^5$ yr. 
By that
time, the disk radius is $R_d\sim$ 10 AU, and the mass loss rate is $\sim 10^{-9}$
\Myr. In this later evolutionary stage, the
small, slowly evaporating disk has a lifetime of $\sim 2 \times 10^6$ yr, 
comparable to the age of SO587 itself.

This scenario can explain our data, as long as SO587 stays in the
vicinity of \sori\ for $\simgreat 10^6$ yr.
With a velocity dispersion of $\sigma \sim 1$ km/s in the radial direction (Sacco et al.,~2008),
 the distance between \sori\ and SO587 can increase to $\sim 1$pc in $10^6$ yr,
i.e., about 3 times the present projected distance. This is still consistent with
the constraints we have set at the beginning of this section, although
marginally so.

However, it is possible that the photoevaporation of the SO587 disk is caused by
the high-energy radiation from the central star.
Recently, Gorti \& Hollenbach,~(2008)
model the case of a low mass TTS (0.3 \Msun), 
with properties roughly comparable to those of SO587. 
The trend of \Mloss\ with time is similar, with
an early phase when the disk is large and \Mloss$\simgreat 10^{-8}$ \Myr,
followed by a decrease in the disk radius and of \Mloss. The phase with
\Mloss $\sim 10^{-9}$ \Myr, $R_d \sim 30$AU has a lifetime
of about $5 \times 10^6$ yr, which can account for the SO587 observed properties.

In both cases, 
a first period of rapid evaporation, which  produces
a significant reduction of the disk size, is followed by a 
phase of slower evolution, with correspondingly lower values of
\Mloss. We think that SO587 is in this late evolutionary stage, because 
its disk has shrunk to  10--30 AU in radius 
and is now in the last, long phase of slow photoevaporation.
SO587 shines in the optical forbidden lines of species, such as 
[SII] and [NII], only because it is at the right distance from \sori\
at the time of our observations.

The exact values of \Mloss, the location of the ionization front,
the dynamics of the gas, and in particular the line profiles are all
the result of the interaction of several complex physical processes,
that need to be explored.
We note, for example,  
that the observed profiles of the optical forbidden lines
require a line-emitting region to be much larger than the disk, which does not seem to obscure
the receding part of the outflow.
Observationally, there are several possible tests. The disk size can be
constrained directly
by spatially resolved observations at millimeter wavelengths, but also indirectly by photometry at far-infrared wavelengths.
Also, a  basic feature of externally photoevaporated disks
is the displacement between the ionization front and the star;
in SO587 this is predicted to be $\sim 0.25$ arcsec,
detectable with spectroastrometric techniques or by direct
imaging from HST.

\section {Discussion and conclusions}


The possibility that we are detecting the final, but long-lasting stages
of the dissipation of a disk by photoevaporation 
is intriguing. 

As mentioned in \S 1, \sori\ is particularly suited to this
kind of study. First of all, the low background emission 
is very favorable for detecting
objects, such as SO587, with a 
mass-loss rate  at least two orders of
magnitude lower than that of the Orion proplyds, and, more generally,
for studying the effects
of an external, high-energy radiation field on 
jets (Andrews et al.,~2004) and disks. 

A second important point is the age of  \sori\
(2--3 Myr),
significantly more than of Orion, such that disks  have had enough time to evolve to the stage of low-accretion
rate, low photoevaporation mass loss. 
We note that \Mloss $\sim$ \Macc $\sim 10^{-9}$ \Myr in SO587, as
expected at time $\simgreat 10^6$ yr in disk evolution models where the photoevaporation
by an external source is coupled to viscous evolution
(Clarke,~2007).

Some disks, such as SO587, show evidence of grain growth and settling in their SEDs, 
and it seems possible not only to detect old photoevaporated disks but also to relate 
their properties to those of the grains.
Larger grains have two competing effects on the 
photoevaporation process: on the one hand, they have lower
extinction in the FUV, allowing the FUV radiation to penetrate
more deeply into the disk; on the other, the heating efficiency
will be lower. 
The balance between these two competing effects
needs to be explored in the model calculations, but
can also be addressed observationally.

Within 0.4 pc around \sori\, Hernandez et al.~(2007) detect about 30 objects 
with all kinds of disks (Class II, Class III, and EV).
The spread of SEDs probably
reflects a varying degree of dust evolution (Dullemond \& Dominik,~2004;
D'Alessio et al.,~2006).
We are currently carrying on  a comprehensive study
of the accretion and mass-loss properties of \sori\ 
objects to identify
other proplyds and to address 
some of the issues mentioned in this section.\\ \\
\begin{small}
Acknowledgments: We would like to thank S. Wolk and D. Hollenbach for useful discussions. 
We also thank the anonymous referee for comments that clarified the paper. 
\end{small}

\bibliographystyle{aa}

\clearpage

\end{document}